\documentclass[letterpaper]{article} % DO NOT CHANGE THIS
\usepackage{aaai25}  % DO NOT CHANGE THIS
\usepackage{times}  % DO NOT CHANGE THIS
\usepackage{helvet}  % DO NOT CHANGE THIS
\usepackage{courier}  % DO NOT CHANGE THIS
\usepackage[hyphens]{url}  % DO NOT CHANGE THIS
\usepackage{graphicx} % DO NOT CHANGE THIS
\usepackage{natbib}  % DO NOT CHANGE THIS AND DO NOT ADD ANY OPTIONS TO IT
\usepackage{caption} % DO NOT CHANGE THIS AND DO NOT ADD ANY OPTIONS TO IT
\usepackage{algorithm}
\usepackage{algorithmic}

\usepackage{booktabs}
\usepackage{threeparttable}
\usepackage{multirow}
\usepackage{makecell}
\usepackage{tabularx}
\usepackage{newtxmath} 
\usepackage{graphicx}
\usepackage{subfigure}
\usepackage{subcaption}
\usepackage{amsmath}
\usepackage{makecell}

\usepackage{newfloat}
\usepackage{listings}
\DeclareCaptionStyle{ruled}{labelfont=normalfont,labelsep=colon,strut=off} % DO NOT CHANGE THIS
\floatstyle{ruled}
\newfloat{listing}{tb}{lst}{}
\floatname{listing}{Listing}
\title{M2I2: Learning Efficient Multi-Agent Communication via Masked State Modeling and Intention Inference}
\author{
    Chuxiong Sun\textsuperscript{\rm 1 \rm 2} \equalcontrib,
    Peng He\textsuperscript{\rm 4 \rm 1} \equalcontrib,
    Qirui Ji\textsuperscript{\rm 1 \rm 3} \equalcontrib,
    Zehua Zang\textsuperscript{\rm 1 \rm 3},\\
    Jiangmeng Li\textsuperscript{\rm 1 \rm 2} \thanks{Corresponding author.},
    Rui Wang\textsuperscript{\rm 1 \rm 2},
    Wei Wang\textsuperscript{\rm 4}
}
\affiliations{
    \textsuperscript{\rm 1}National Key Laboratory of Space Integrated Information System, Institute of Software Chinese Academy of Sciences\\
    \textsuperscript{\rm 2}State Key Laboratory of Intelligent Game\\
    \textsuperscript{\rm 3}University of Chinese Academy of Sciences\\
    \textsuperscript{\rm 4}Beijing University of Posts and Telecommunications
    \{chuxiong2016, jiqirui2022, zehua2020, wangrui, jiangmeng2019\}@iscas.ac.cn, \{hepeng123, weiwang\}@bupt.edu.cn
}

\usepackage{bibentry}
\begin{document}

\maketitle

\begin{abstract}
Communication is essential in coordinating the behaviors of multiple agents. However, existing methods primarily emphasize content, timing, and partners for information sharing, often neglecting the critical aspect of integrating shared information. This gap can significantly impact agents' ability to understand and respond to complex, uncertain interactions, thus affecting overall communication efficiency. To address this issue, we introduce M2I2, a novel framework designed to enhance the agents' capabilities to assimilate and utilize received information effectively. M2I2 equips agents with advanced capabilities for masked state modeling and joint-action prediction, enriching their perception of environmental uncertainties and facilitating the anticipation of teammates' intentions. This approach ensures that agents are furnished with both comprehensive and relevant information, bolstering more informed and synergistic behaviors. Moreover, we propose a Dimensional Rational Network, innovatively trained via a meta-learning paradigm, to identify the importance of dimensional pieces of information, evaluating their contributions to decision-making and auxiliary tasks. Then, we implement an importance-based heuristic for selective information masking and sharing. This strategy optimizes the efficiency of masked state modeling and the rationale behind information sharing. We evaluate M2I2 across diverse multi-agent tasks, the results demonstrate its superior performance, efficiency, and generalization capabilities, over existing state-of-the-art methods in various complex scenarios. 
\end{abstract}

\section{Introduction}
Reinforcement Learning (RL) has achieved significant milestones in various complex real-world applications, from Game AI \cite{osband2016deep,silver2017mastering,silver2018general,vinyals2019grandmaster} and Robotics \cite{robotics} to Autonomous Driving \cite{leurent2018survey}. However, the landscape shifts markedly when applied to Multi-Agent Reinforcement Learning (MARL) \cite{maddpg,qmix1,mappo1}, where unique challenges emerge. A principal challenge is the issue of partial observability, where agents must make decisions based on limited local observations, lacking a comprehensive view of the entire environment.  
In addressing this challenge, multi-agent communication emerges as a potent solution. By enabling agents to \emph{share} and \emph{integrate} information, this strategy facilitates a deeper collective understanding of their environment, stabilizing the learning process and promoting synchronized actions among agents.   

Despite advancements, existing methods in multi-agent communication primarily focus on sending policies, such as creating meaningful messages \cite{vbc,tmc,maic}, optimizing the timing \cite{ic3net,schednet} and selecting appropriate partners \cite{I2C1,magic1,sms} for information exchange. However, these methods exhibit a significant gap in effectively integrating received information to enhance decision-making at receiving end. Typically, a large volume of received messages, processed by basic mechanisms like concatenation \cite{commnet111} is fed directly into policy networks. This approach treats the information integration task as a black box, presupposing that neural networks can autonomously discern the most decision-important information, overlooking the intricacies of cognition and collaborative decision-making. In contrast, human cognitive processes \cite{cog} demonstrate a superior ability for utilizing received information to perceive the environment and reduce the uncertainty of decision-making. 
This level of decision-making complexity, inherent in human cognition, is something that current multi-agent communication methods fail to capture.

Inspired by human cognitive processes and recent advancements in representation learning, as illustrated by models like BERT \cite{bert} and MAE \cite{mae}, we redefine the challenge of information integration in multi-agent communication as one of representation learning task. 
In this context, agents are tasked with developing representations for cooperative decision-making from a limited set of messages. These messages, constrained by partial observability and limited communication resources, only reflect a subset of the environmental states, often proving inadequate for a comprehensive understanding of environmental dynamics. Furthermore, not all received messages are beneficial for decision-making; some may introduce noise that disrupts the process. Consequently, we argue that an ideal representation in this context must be both \textbf{sufficient}—offering a comprehensive breadth of information for a deep understanding of the environment, and \textbf{informative}—sharply focused on data crucial for facilitating cooperative decision-making.

Following this principle, we introduce M2I2, a novel approach incorporating two self-supervised auxiliary tasks to enhance efficiency of information integration.  %and explore intrinsic rationale of information sharing. 
To meet the standard of ``sufficient", M2I2 utilizes masked modeling techniques to reconstruct global states from received messages, furnishing agents with comprehensive information for informed decision-making. Essentially, M2I2 introduces a state-level Masked Auto-Encoder (MAE) designed for multi-agent communication. A distinctive aspect of this model is its unique masking mechanism, where the masks are dynamically determined by the communication strategies of the sending agents. Our empirical studies highlight that traditional random mask generating techniques \cite{bert,mae,maskdp} fall short in addressing the complexities encountered in MARL. 
%A distinctive feature of M2I2's masked state modeling is its adaptive masking process, which responds to both environmental constraints and the strategies of sending agents. The reason behind the behaviour of this nuanced approach is that our empirical studies reveal that traditional random masking techniques \cite{bert,mae,maskdp} are insufficient for the masked state modeling due to the complexities of MARL environments. 
% To this end, we develop the Dimensional Rational Network (DRN), trained via a meta-learning paradigm, which dynamically adjusts the importance of each dimension of observed information, taking into account their impact on both decision-making and auxiliary tasks. 
To this end, we develop the Dimensional Rational Network (DRN) to dynamically adjust the importance of each dimension of observed information. DRN is trained via a meta-learning paradigm, which takes into account the impacts on both decision-making and auxiliary tasks.
After exploring the rationale of dimensional observations, we further propose an importance-based heuristic to discern which dimensions of observations should be masked at both training and execution stages, thereby enhancing the efficiency of masked state modeling and communication rationality.

Regarding the ``informative" aspect, M2I2 integrates an inverse model to predict joint actions from sequential state representations, enabling agents to focus on information pivotal to their decisions. Furthermore, the inverse model enables agents to infer their teammates' intentions during decentralized decision-making processes.
% Such a capability is crucial for ensuring that communication within the team is not only about sharing information but also about understanding intentions and insights of their teammates that can influence collective strategies.
This capability is essential for facilitating team communication that goes beyond mere information exchange, enabling a deeper understanding of teammates' intentions and insights that can impact collective strategies.
By introducing the self-supervised objective, M2I2 facilitates a deeper integration of received information, allowing agents to align closely with each other's intentions and leading to more efficient and informed decision-making across various scenarios.
%It is important to note that the proposed auxiliary tasks are used solely during the training phase. Their purpose is to enhance the agents' encoders, enabling them to generate sufficient and informative representations based on limited observations during the execution phase, in line with the Centralized Training with Decentralized Execution (CTDE) paradigm.
%A distinctive feature of M2I2's masked state modeling is its adaptive masking process, which responds to both environmental constraints and the strategies of sending agents. The reason behind the behaviour of this nuanced approach is that our empirical studies reveal that traditional random masking techniques \cite{bert,mae,maskdp} are insufficient for the masked state modeling due to the complexities of MARL environments. To this end, we develop the Dimensional Rational Network (DRN), trained via a meta-learning paradigm, which dynamically adjusts the importance of each dimension of observed information, taking into account their impact on both decision-making and auxiliary tasks. After exploring the rationale of dimensional observations, we further propose an importance-based heuristic to discern which dimensions of observations should be masked at both training and execution stages. This approach significantly enhances the efficiency of masked state modeling and communication rationality.

To validate the effectiveness of M2I2, we conduct comprehensive evaluations across a range of multi-agent tasks with differing complexities, from Hallway and MPE to SMAC. Compared to state-of-the-art communication methods \cite{tarmac,maic,sms,masia}, M2I2 demonstrates superior performance, enhanced efficiency, and remarkable generalization capabilities. Our main \textbf{contributions} are summarized in three-fold:
\begin{itemize}
    \item To the best of our knowledge, M2I2 represents the first instance of incorporating self-supervised objectives, i.e., reconstructing global states and predicting joint actions, into the process of information integration under the condition of partial observability and restricted communication resources.
    \item We integrate a meta-learning paradigm to model the contribution of each dimensional piece of information towards both decision-making and self-supervised objectives, therefore directing agents to transmit and focus on only the most relevant and important information.
    \item  Empirically, our proposed method not only facilitates efficient message integration, but also significantly improves communication efficiency, effectively bridging a vital research gap in MARL.
\end{itemize}

\section{Related Works}%compare to tom2c and masia
Multi-agent communication has emerged as an indispensable component in MARL. %, particularly in addressing the challenges of partial observability and complex agent interactions. 
Research in this domain has primarily concentrated on three fundamental questions:  

\textbf{\emph{Determining the optimal content of communication (what to communicate).}} 
CommNet \cite{commnet111}, as a pioneering work in this area, facilitated agents in learning continuous messages. Following CommNet, several methods have been developed to further refine the message learning process. VBC \cite{vbc} aims to filter out noisy parts while retaining valuable content by limiting the variance of messages. TMC \cite{tmc} introduces regularizers to reduce temporally redundant messages. NDQ \cite{ndq} employs information-theoretic regularizers to develop expressive and succinct messages. MAIC \cite{maic} enabled agents to customize communications for specific recipients, advancing tailored message learning.

\textbf{\emph{Deciding appropriate timing and partners for information exchange (when and whom to communicate).}}
To enhance communication efficiency, approaches such as IC3Net \cite{ic3net} and ATOC \cite{schednet} have introduced gating networks  to eliminate superfluous communication links. Similarly, SchedNet \cite{schednet} and IMMAC \cite{immac} have modeled the significance of observations, using heuristic mechanisms to gate non-essential communication.  Further, methods such as MAGIC \cite{magic1}, I2C \cite{I2C1} and SMS \cite{sms} have been developed to identify the most suitable recipients. 
These approaches focus on modeling the contribution of shared information to the decision-making processes of the recipients, aiming to direct communication where it most influences decision-making.

\textbf{\emph{Integrating incoming messages and making decisions (how to utilize received information).}} 
%However, a significant focus of these methods has been on the transmission of information, with less attention to the critical aspect of information integration at the receiver's end. Only a few studies 
TarMAC \cite{tarmac} has explored how agents can effectively assimilate crucial information from an abundance of raw messages. MASIA \cite{masia} take a different approach, employing an Auto-Encoder and a forward model for information integration and becoming the first to introduce self-supervised learning into multi-agent communication.  However, MASIA is under the strong assumption that agents have access to all observations from their peers. In this work, we challenge and relax this assumption by introducing the masked state modeling technique, extending the approach to more realistic environments where communication resources are constrained.  %Yet, it operates under broadcast and full communication assumptions, creating a disconnect with practical application scenarios. 

Furthermore, our work is also related to the mask modeling techniques \cite{mae, bert, maskdp}. However, M2I2 is the first to apply this approach within the multi-agent communication domain, utilizing this technique to effectively tackle the challenge of imperfect information that arises from constrained communication resources. A key distinction of M2I2 lies in its innovative masking mechanism, which differs significantly from those used in prior methods. This unique approach will be elaborated upon in \textbf{Section \ref{drn}}.
% To the best of our knowledge, this work represents the first instance of incorporating self-supervised objectives—specifically, reconstructing global states and predicting joint actions—into the process of information integration under the condition of restricted communication. Moreover, we integrate a meta-learning paradigm to precisely model the contribution of each dimensional piece of information towards both decision-making and self-supervised objectives, therefore directing agents to generate and transmit only the most relevant and crucial information. This novel approach not only facilitates efficient message integration but also significantly improves communication efficiency, effectively bridging a vital research gap in MARL.

%\subsection{Self-Supervised Learning}

\section{Preliminary}
%\subsection{Decentralized Partially Observable Markov Decision Processes (Dec-POMDPs)}
In this work, we focus on fully cooperative multi-agent tasks, characterized by partial observability and necessity for inter-agent communication. These tasks are modeled as Decentralized Partially Observable Markov Decision Processes (Dec-POMDPs) \cite{decpomdp}, represented by the tuple $G = (N, S, O, A, \mathbb{O}, P, R, \gamma, M)$. In this formulation, $N \equiv \{1, \ldots, n\}$ denotes the set of agents, $S$ represents the global states, $O$ describes the observations available to each agent, $A$ signifies the set of available actions, $\mathbb{O}$ is the observation function mapping states to observations, $P$ is the transition function illustrating the dynamics of the environment, $R$ is the reward function dependent on the global states and joint actions of the agents, $\gamma$ is the discount factor, and $M$ specifies the set of messages that can be communicated among the agents. 
At each time-step $t$, each agent $i \in N$ has access to its own observation $o_i^t \in O$ determined by the observation function $\mathbb{O}(o_i^t | s_t)$. Additionally, each agent can receive messages $c_i^t = \sum_{j \neq i}m_j^t$ from teammates $j \in N$. Utilizing both the observed and received information, agents then make local decisions. As each agent selects an action, the joint action $a_t$ results in a shared reward $r_t = R(s_t, a_t)$ and transitions the system to the next state $s_{t+1}$ according to the transition function $P(s_{t+1}|s_t,a_t)$. The objective for all agents is to collaboratively develop a joint policy $\pi$ to maximize the discounted cumulative return $\sum_{t=0}^{\infty}\gamma^{i}r_{t}$.

\begin{figure*}[t]
  \centering
  \includegraphics[width=0.8\textwidth]{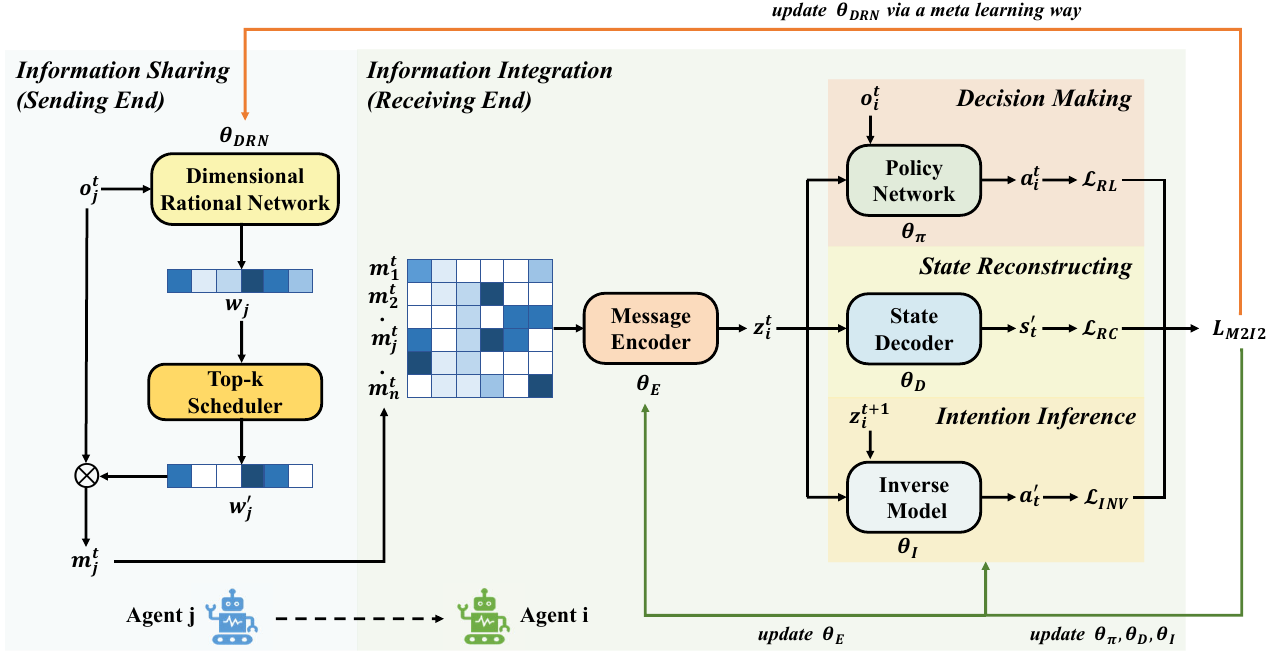}
  \caption{Framework of M2I2. Similar to other CTDE  approaches in MARL, M2I2 only leverages global states and joint actions during centralized training phase. However, M2I2 distinguishes itself through its self-supervised auxiliary tasks. These tasks enable agents to develop representations from received messages, enhancing their ability to comprehend global states and infer teammates' joint actions. This capability becomes particularly valuable during the decentralized execution phase, where agents must operate based on limited observations.
  }
  \label{fig:M2I2}
\end{figure*}

%可以加一段MARL sampling和training的描述。或者是CTDE的描述，然后写一个marl的loss
%\subsection{Centralized Training with Decentralized Execution (CTDE)} 
Centralized Training with Decentralized Execution (CTDE) \cite{maddpg} stands as a promising paradigm in cooperative multi-agent tasks.
Within this paradigm, individual agents make decisions based on local information, while their policies are trained through a centralized manager with access to global information. This work aligns with the prevailing value-based approaches within the CTDE framework. During the training phase, the joint action-value function $Q_{tot}(s_t, a_t; \boldsymbol{\theta})$ will be trained to minimize the expected Temporal Difference (TD) error:
\begin{equation}
    \mathcal{L}_{RL}(\boldsymbol{\theta})
    =\mathbb{E}_{s_t,a_t,r_t,s_{t+1}\in D}\left[y_t-Q_{tot}(s_t,a_t;\boldsymbol{\theta})\right]^2,
\end{equation}
where $D$ is the replay buffer, $y_t = r_t+\gamma max_{a_{t+1}}Q_{tot}(s_{t+1},\\a_{t+1};\boldsymbol{\theta}^-)$ is the TD target. $\boldsymbol{\theta}^-$ denotes the parameters of the target network, which is periodically updated by $\boldsymbol{\theta}$. 
% For a comprehensive understanding of the foundational concepts and specific notations used in this study, please refer to the Appendix section \ref{app:notations}. 

\section{Methodology}
\label{section:method}
%In the following sections, we provide an in-depth exploration of the M2I2 framework, highlighting its key components. We begin with a comprehensive overview of the M2I2's communication structure, offering insights into its design and functionality. Then, we shift our focus to two critical auxiliary tasks formulated within M2I2. Finally, we delve into the efficiency of training the Dimensional Rational Network (DRN). 

%we introduce the Dimensional Rational Network (DRN), a novel framework designed to meticulously assess and underscore the significance of observed information at the dimensional level. This network plays a crucial role in fine-tuning the information filtering process, ensuring relevance and precision in communication.  

\subsection{Overall Framework} \label{framework}
The framework of M2I2 are shown in Figure \ref{fig:M2I2}, with its core components highlighted for effective multi-agent communication. A key component of M2I2 includes a message encoder and a state decoder, functioning collectively as an extendable module to reconstruct the environmental states in an auto-encoding manner. This design allows for the reconstruction of global states from received messages (i.e. limited observations), thereby providing agents with sufficient information to make well-informed cooperative decisions. %This feature is instrumental in overcoming the challenges posed by partial observability and limited communication resources. 
Furthermore, M2I2 integrates an inverse model capable of predicting joint actions based on consecutive state representations. This model is pivotal in equipping agents with the ability to infer the intentions of their teammates while making decisions. 
%Such a capability is crucial for enhancing the effectiveness of cooperative strategies and interactions among agents, as it allows for a more nuanced understanding and anticipation of team dynamics and coordination. 
Another standout feature of M2I2 is DRN, which is adept at evaluating the importance of various observed information based on their gradient contributions to both auxiliary and RL tasks.  The DRN is continually refined through a meta-learning paradigm, which effectively avoid the trivial solution and local optimum issues during training. By identifying and emphasizing important information, DRN enables agents to share and focus on important data, thereby optimizing the communication process for efficiency and effectiveness. 

\subsection{Communication Process of M2I2}
% Specifically, 
The communication process of M2I2 can be summarized as the following four steps.

\textbf{Selectively masking unnecessary observations for information sharing}: At each time-step, agents utilize the DRN to evaluate the importance of observed information in supporting decision-making and auxiliary tasks. This importance is quantified as $\mathcal{\omega}_i=\mathcal{\omega}_{id}|d\in[1, D]$, where $i$ represents the ID of the agent and $D$ is the dimensionality of observed information. 
To optimize communication efficiency while ensuring effective decision-making, %and completion of auxiliary tasks, 
a topK mechanism is applied for generating observation masks, which is formulated as:
\begin{equation}
    \text{topK}(\omega_i)=\begin{cases} 
        \omega_{id}, & \text{if } d \text{ in top-k largest dimensions}  \\
        0, & \text{others}.
    \end{cases}
\label{topk}
\end{equation}
This process allows each agent to selectively share the most important dimensions of the observations while non-essential dimensions are masked to zero. 
The resulting shared information is represented as:
\begin{equation}
m_i^t=o_i^t \otimes \text{topK}(\mathcal{\omega}_i),
\label{message_generate}
\end{equation}
where $m_i^t$ denotes the messages, i.e. masked and weighted observations, and $\otimes$ is an element-wise Hadamard product function. 
The DRN, central to this selectively observation mask process, is trained using a meta-learning paradigm (detailed in \textbf{Section \ref{drn}}), which enables it to dynamically adjust its assessments based on both decision-making and auxiliary task performances.
% This training approach ensures that the DRN effectively identifies which data dimensions are most impactful, thereby enhancing the overall communication strategy within the multi-agent system. 
% For a detailed discussion on the functionalities and training of the DRN, please refer to \textbf{Section} \ref{drn}.

\textbf{Integrating received information}: Upon receiving messages, M2I2 integrates a scaled dot-product self-attention module \cite{attention} to adeptly process incoming messages. Specifically, the received messages are transformed into corresponding queries $Q$, keys $K$, and values $V$. The process of integrating this information is mathematically represented as follows:
\begin{equation}
    z_i^t = f_{\theta_{E}}(\operatorname{softmax}(\frac{QK^T}{\sqrt{D_k}})V)
\label{intergrated_representation}
\end{equation}
where $\theta_{E}$ represents the parameters of Message Encoder and $D_k$ represents the dimension of a single key. This message encoder exhibits two notable benefits. Firstly, the encoder's design makes it adaptable to diverse communication contexts, accommodating varying numbers and arrangements of agents. Secondly, by utilizing a weighted sum mechanism, the self-attention module integrates information without excessively expanding the agents' local policy spaces. 

\textbf{Implicitly Inferring the global states and teammates' intention}: Following this, M2I2 encode the received messages into a compact representation. Unlike traditional methods that rely solely on RL objectives, which often struggle to learn effective representation from the limited and noisy messages, M2I2 incorporates two self-supervised objectives. These objectives are specifically designed to  develop a representation that is both ``sufficient" for a thorough understanding of the environment and ``promising" for aiding cooperative decision-making. 
\emph{Although the involved self-supervised auxiliary tasks are conducted only during training, they can implicitly enhance the message encoder's ability to interpret the environment and predict teammates' intentions during decentralized decision-making process.}
% For an in-depth exploration of these auxiliary tasks and their functions, please refer to \textbf{Section} \ref{section:aux}. 

\textbf{Making cooperative decisions}: The culmination of the M2I2 process is reflected in the agents' ability to make cooperative decisions. Here, the enriched integrated information, blended with each agent's personal observations, is channeled into the policy network. The process of decision-making is mathematically represented as follows:
\begin{equation}
    a_i^t = \pi_i(o_i^t, z_i^t;\theta_{\pi})
\label{action_select}
\end{equation}
where $\theta_{\pi}$ represents the parameters of policy network. This convergence of individual perception and collective insights is crucial, as it empowers agents to make decisions that are not only informed, but also aligned with the overarching goals and strategies of the team.

\subsection{Self-Supervised Auxiliary Tasks for Efficient Multi-Agent Communication}
\label{section:aux}

% We provide an in-depth exploration of auxiliary tasks proposed by M2I2. 
Given the inherent constraints in agents' perceptual capabilities and the limitations of communication bandwidth, the information encoded by the message encoder often captures only a fraction of the environment's state. To ensure that agents have access to sufficient information for effective decision-making, we employ a state decoder. This decoder is tasked with reconstructing the global state of the environment, represented by $s_t^{\prime}=g_{\theta_{D}}(z_i^t)$. The associated loss function is computed using the mean squared error between the reconstructed and global states:
\begin{equation}
    \mathcal{L}_{RC}(\theta_{E},\theta_{D})=\mathbb{E}_{ z_i^t,s_t}\|s_t^{\prime}-s_t\|_2^2.
\label{lossRC}
\end{equation}
By combining the message encoder with the state decoder, we effectively create an extendable masked state modeling. This masked modeling is characterized by a unique masking process, generated both by the environment and the agents themselves. This approach enables the integrated representation $z_t$ to effectively represent the global states of the environment, thus overcoming the challenges posed by their limited observational scope and communication capacity.

To augment the capability of agents in focusing on information promising to their decisions and aligning with the intentions of their counterparts, we introduce an inverse model, denoted as $I_{\theta_{I}}:\overset{\cdot}{\mathcal{Z}\times\mathcal{Z}}\to A^{n}$, where $\mathcal{Z}$ is the space of state representations. This model is crafted to predict the joint actions that agents take to transition from one state representation to the next. Formally, given a triplet $(z_t,a_t,z_{t+1})$ composed of two consecutive state representations and joint actions taken by agents, we parameterise the conditional likelihood as $p(a_t^{\prime}) = I_{\theta_{I}}(z_t,z_{t+1})$, where $I_{\theta_{I}}$ embodies a two hidden layers MLP followed by a softmax operation. The parameters of both inverse model $\theta_{I}$ and message encoder $\theta_{E}$ are optimized via a maximum likelihood approach. The corresponding loss function is formulated as: 
\begin{equation}
    \mathcal{L}_{INV} (\theta_{E},\theta_{I}) = \mathbb{E}_{ z_i^t,a_t}\|a_t^{\prime}-a_t\|_2^2,
\label{lossINV}
\end{equation}
where this loss function measures the discrepancy between the predicted joint actions and the actual joint actions. At first, the objective encourages agents to focus on information that are controllable and expressive pertinent to cooperative decision-making. This focus is crucial for agents to effectively handle elements that they can influence, enhancing their relevance in a coordinated environment. Moreover, the deeper integration of received information facilitated by the model allows agents to implicitly infer the intentions of others during execution.  This capability significantly bolsters agents' potential to align their actions with the intentions of their teammates. Such alignment is not only technically beneficial, but also aligns with cognitive research findings \cite{cog}, which underscore the importance of intention understanding in effective social interactions.

% \subsection{Training Scheme}
\subsection{DRN for Importance Modeling} \label{drn}
% In the framework of M2I2, the DRN plays a pivotal role. 
DRN is designed to discern the importance of different dimensions of observed information, specifically tailoring to the needs of  decision-making and auxiliary tasks. The primary challenge here lies in the dynamic nature of the MARL and the variability in communication needs across different stages of a mission. Unlike static scenarios, the importance of information can change dramatically, requiring the DRN to adapt continuously and efficiently. This challenge transcends the realm of simple optimization problems, typically addressed with first-order gradients. 
% To effectively navigate this complexity, we employ a meta-learning approach \cite{meta}.  
% The meta-learning paradigm is suited because it can circumvent trivial solutions and local optima that often hinder training efficiency. 
To effectively navigate this complexity, we employ a meta-learning approach \cite{meta}, as it is well-suited to circumvent trivial solutions and local optima that often hinder training efficiency.
It allows the DRN to dynamically adjust its understanding of information importance in a sophisticated manner, aligning closely with the overarching goals of decision-making and auxiliary tasks. 

It is important to note that within our training framework, only the parameters of $\theta_{DRN}$ are refined using this meta-learning approach. The other parameters of the system are updated using conventional first-order gradient methods. %Specially, the training within the M2I2 framework is a sophisticated process, involving a meta-learning approach that can be unfolded into two distinctive steps. This intricate training regimen is designed to ensure that DRN finely tunes its understanding of the importance of various dimensions in the context of the tasks at hand. 
Specifically, in the first regular training step, we focus on training the combined set of parameters $\theta = (\theta_{E}, \theta_{D}, \theta_{I}, \theta_{\pi})$ by jointly minimizing the auxiliary tasks and RL losses, which is formalized by
\begin{equation}
    \arg\min_{\theta}\mathcal{L}_{M2I2}(\theta, \theta_{DRN}), 
\label{lossM2I2}
\end{equation}
where $\mathcal{L}_{M2I2}(\theta, \theta_{DRN}) = \mathcal{L}_{RL} + \beta (\mathcal{L}_{RC} + \mathcal{L}_{INV})$ and $\beta$ is a coefficient that controls the balance between RL objective and auxiliary objectives.

% \begin{figure*}[t]
% \centering
% \subfigure{
% \begin{minipage}[t]{0.15\textwidth}
% \centering
% \includegraphics[width=1.0\linewidth]{hw_state(1).pdf}
% \subcaption*{(a) Hallway}
% \end{minipage}%
% }%
% \quad
% \subfigure{
% \begin{minipage}[t]{0.15\textwidth}
% \centering
%  \includegraphics[width=1.0\textwidth]{pradator(1).pdf}
% \subcaption*{(b) Predator Prey}
% \end{minipage}%
% }%
% \quad 
% \subfigure{
% \begin{minipage}[t]{0.25\textwidth}
% \centering
% \includegraphics[width=1.0\linewidth]{smac.png}
% \subcaption*{(c) SMAC}
% \end{minipage}%
% }%
% \quad 
% \subfigure{
% \begin{minipage}[t]{0.25\textwidth}
% \centering
% \includegraphics[width=1.0\linewidth]{smac-communication.png}
% \subcaption*{(c) SMAC-Communication}
% \end{minipage}
% }%
% \caption{Multiple environments considered in our experiments.}
% \label{scenarios}
% \end{figure*}

In the second meta-learning-based step, $\theta_{DRN}$ is updated by using the second-derivative technique \cite{meta,DBLP:conf/icml/LiQZ0X22}. 
This technique is crucial for adjusting $\theta_{DRN}$ to better discern the importance of various information dimensions that significantly impact both RL and auxiliary tasks. 
%The objective of this phase is to steer $\theta_{DRN}$ towards recognizing the significance of dimensions that are relevant or irrelevant to the RL and auxiliary tasks. 
%These dimensions, regarded as dimensional confounders, can obscure the encoders' ability to effectively isolate and emphasize discriminative information crucial for both sampling and training processes. 
The update process involves calculating the gradients of $\theta_{DRN}$ in relation to the combined performance metrics from these tasks, encapsulated by $\mathcal{L}_{M2I2}$. 
%To effectively update $\theta_{DRN}$, we calculate its gradients in relation to the performance metrics of both the auxiliary and RL tasks, i.e., via $\mathcal{L}_{M2I2}$. 
% This process involves utilizing the gradients obtained during the back-propagation of the comprehensive loss function, $\mathcal{L}_{M2I2}$. 
Formally, we update $\theta_{DRN}$ by 
\begin{equation}
    \arg\min_{\theta_{DRN}}\mathcal{L}_{M2I2}(\theta_{trial}, \theta_{DRN}),
\label{updatemeta}
\end{equation}
where $\theta_{trial}=(\theta_{E}^{trial}, \theta_{D}^{trial}, \theta_{I}^{trial}, \theta_{\pi}^{trial})$ is the trial weights of the $\theta$ after one gradient update using the M2I2 loss defined in Equation \ref{lossM2I2}. 
We formulate the updating of such trial weights as follows:
\begin{equation}
\theta_{trial}=\theta-\ell_\theta\nabla_\theta\mathcal{L}_{M2I2},  
\label{trailcompute}
\end{equation}
where $\ell_\theta$ is the learning rate.
Note that the calculation of trial weights excludes the step of gradient back-propagation.
Thus, $\theta_{DRN}$ is updated through the second-derivative gradient of $\theta$.
By doing so, we ensure that $\theta_{DRN}$ is continuously fine-tuned by the gradient contributions of $\mathcal{L}_{M2I2}$, allowing DRN to dynamically evaluate the importance of each observed dimension. % by assessing how changes in these dimensions impact the overall performance.   
Our visualization study in \textbf{Appendix \ref{app:visualization}} further validates this approach: it reveals that the DRN not only distinguishes varying levels of importance across different agent types and observation categories but also dynamically adjusts these importance weights over time, adapting to the evolving demands of the task.

% The underlying principle of our method involves performing a derivative of the derivative (a Hessian matrix computation) with respect to the combined parameter set $\theta$ to update $\theta_{DRN}$. More specifically, this process entails computing the derivative relative to $\theta_{DRN}$ utilizing a retained computational graph of $\theta$. Subsequently, $\theta_{DRN}$ is updated through the back-propagation of this derivative, as outlined in Equation \ref{updatemeta}. 
% This procedure is characterized by a cyclical nature, where the steps for updating $\theta$ and $\theta_{DRN}$ are iteratively carried out. This iterative process ensures that the updates to $\theta_{DRN}$ are informed by and in turn influence the broader learning framework encapsulated by $\theta$.  

%explain why utilize meta to capture imporatance of observations.

\section{Experiment}
\label{Experiment}
In this section, our experimental design is meticulously structured to address three fundamental questions:
\begin{itemize}
    \item \textbf{RQ1.} How does M2I2's performance and efficiency compare to leading communication methods?
    \item \textbf{RQ2.} What specific components within M2I2 are instrumental to its performance?
    \item \textbf{RQ3.} Is M2I2 versatile enough to be applied across a range of tasks, and can it be seamlessly integrated with multiple existing baselines?
\end{itemize}

\subsection{Setup}
\label{Experiment_setup}

\textbf{Benchmarks.}
\label{Experiment_benchmark}
In order to demonstrate the effectiveness and generality of M2I2, we conducted extensive experiments across four popular multi-agent communication benchmarks: Hallway \cite{ndq}, Predator-Prey (PP) \cite{maddpg}, SMAC \cite{smac} and SMAC-Communication \cite{ndq}. Each of these benchmarks provides a substantial testbed for evaluating multi-agent communication strategies. Detailed descriptions of each environment can be found in \textbf{Appendix} \ref{app:experimentalsetup}.

\begin{figure*}[t]
   \centering
   \includegraphics[width=1.0\textwidth]{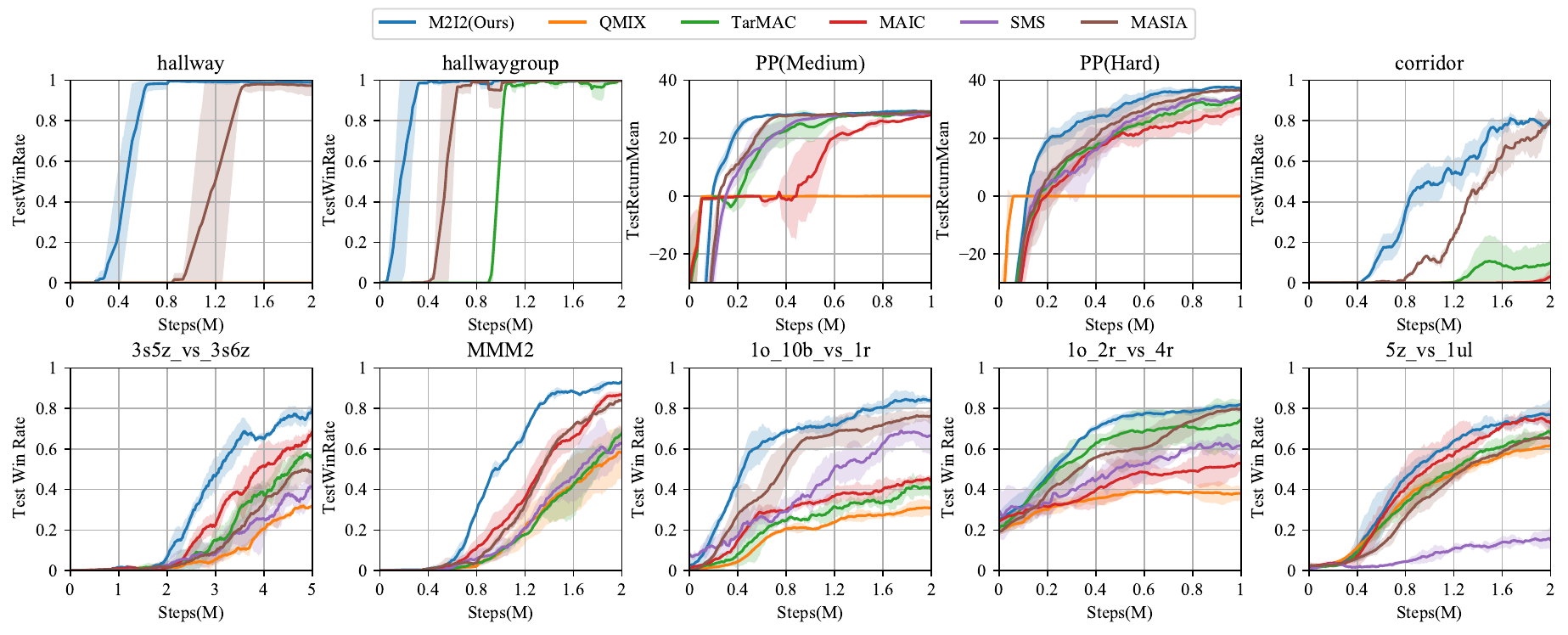}
   \caption{Performance on multiple benchmarks.
   }
   \label{performance}
 \end{figure*}

\textbf{Baselines.}
For comparative analysis, we select a diverse set of baselines. This includes QMIX \cite{qmix1}, a well-established MARL algorithm that operates without a communication mechanism. To assess our method's performance in the context of communication-enhanced MARL, we also include contemporary state-of-the-art communication methods: TarMAC \cite{tarmac}, MAIC \cite{maic}, SMS \cite{sms}, and MASIA \cite{masia}. Each of these baselines represents a significant stride in the development of communication strategies within the MARL framework, providing a robust backdrop for evaluating the efficacy and innovation of our proposed approach. 

\textbf{Hyperparameters}. To ensure reproducibility, the intricate details of our method’s architecture, and our hyperparameter choices are extensively detailed in the \textbf{Appendix} \ref{app:experimentalsetup}.

\subsection{Performance (RQ1)}
\label{Experiment_performance}
Our evaluation begins with a comparative analysis of the learning curves of M2I2 against a range of baseline methods across diverse environments. This comparison is aimed at assessing the comprehensive performance of M2I2. As depicted in Figure \ref{performance}, M2I2 demonstrates a notable performance advantage, consistently outperforming all baselines by a significant margin in each tested environment, indicating M2I2’s strong applicability across scenarios of varying complexity. Specifically, in Hallway, where the reward signals of environment is sparse, many methods exhibit poor performance or fail to learn effectively. In contrast, M2I2 rapidly achieves a 100\% win rate. This success can be attributed to our proposed auxiliary tasks, which appear to significantly aid agents in understanding the locations and intentions of their teammates. In PP, where the communication-free method QMIX struggles, most communication methods demonstrate effectiveness. Notably, M2I2 achieves the best sample efficiency, swiftly identifying the optimal policy.  In SMAC and SMAC-Communication, several full communication methods face challenges in scenarios with large joint observation spaces. For instance, TarMAC shows competence in  $1o\_2r\_vs\_4r$ and $5z\_vs\_1ul$ but underperforms in $corridor$ and $1o\_10b\_vs\_1r$. This could be attributed to the naive approach of these methods in feeding observations from all agents directly into the policy network, thereby increasing the complexity of policy learning. In contract, M2I2 consistently excels in all six SMAC scenarios, irrespective of the varying difficulties, number of agents, and terrain types. This further underscores the broad applicability and potency of M2I2 in diverse multi-agent communication contexts.

%十个曲线图 所有场景的性能对比，scenarios: Hallway(4 6 8 10), Hallway_Group, Predator-Prey 15*15, Predator-Prey 20*20, corridor, 3s5z_vs_3s6z, MMM2, 5z_vs_1ul, 1o_10b_vs_1r, 1o_2r_vs_4r. Baseline: QMIX DOP MAPPO(备选) SMS TARMAC MAIC MASIA 

\begin{table}[t]
\centering
\small
\setlength{\tabcolsep}{5pt}
\begin{tabular}{c|cccc}
  \toprule
  Communication & \multirow{2}*{Hallway} & \multirow{2}*{PP} & \multirow{2}*{SMAC} & SMAC-\\
  Efficiency & ~ &&& Communication \\
  \hline
  TarMAC & 49.6\% & 32.32 & 19.1\% & 17.4\%\\
  MAIC & 0.0\% & 29.75 & 27.1\% & 10.1\%\\
  SMS & 0.0\%  & 51.63 & 13.7\% & 41.8\% \\
  MASIA & 98.6\% & 32.52 & 46.5\% & 28.5\% \\\hline
  M2I2(Ours) & \textbf{165.6\%}& \textbf{55.76} & \textbf{98.7\%}  & \textbf{59.3\%}  \\
  \bottomrule
\end{tabular}
\caption{Communication Efficiency}
\label{table:efficiency}
\end{table}

\subsection{Efficiency (RQ1)}

%一个表格 包括不同算法在不同场景下的communication rate， performance improvement and communication efficiency Communication rate in different environments, hallway differs from SMAC and PP
Efficiency is a long-standing issue in multi-agent communication, as many real-world applications operate under limited communication resources. Therefore, it is crucial to achieve promising performance while maintaining a low communication resource cost. 
Notably, the performance of M2I2, as reported in Figure \ref{performance}, was achieved with a 60\% communication frequency, where 60\% is a hyper parameter defined by our proposed top-k mechanism in \textbf{Section \ref{framework}}. To further understand the communication efficiency of M2I2, we adopted a mechanism inspired by MAGIC \cite{magic1} to measure communication efficiency. Specifically, we calculated the performance improvement for each communication algorithm by subtracting the baseline performance of their communication-free versions. For the SMS algorithm, the communication-free baseline used was DOP \cite{dop}, while for other algorithms, QMIX served as the baseline. 
Subsequently, we examined the communication frequency for each method. M2I2 and SMS both operated at approximately 60\% communication frequency, in contrast to other methods which utilized 100\% communication frequency. Finally, we calculated the communication efficiency for each method by dividing the performance improvement by the communication frequency. 
As indicated in Table \ref{table:efficiency}, M2I2 demonstrated a substantial lead in communication efficiency across all tested scenarios, further validating its effectiveness. This analysis not only underscores M2I2’s ability to maintain high performance with reduced communication demands but also highlights its significant advantages in terms of resource efficiency.

% \begin{table*}[t]
% \centering
% \begin{tabular}{l|llll}
%   \toprule
%   \makecell[c]{Communication Efficiency} & \thead{Hallway} & \thead{PP} & \thead{SMAC} & \thead{SMAC-Communication}\\\hline
%   \makecell[c]{TarMAC} &   \makecell[c]{49.6\%}   &   \makecell[c]{32.32}    &   \makecell[c]{19.1\%}  &  \makecell[c]{17.4\%}\\
%   \makecell[c]{MAIC}&   \makecell[c]{0.0\%}  &  \makecell[c]{29.75}     &   \makecell[c]{27.1\%}  &  \makecell[c]{10.1\%}\\
%   \makecell[c]{SMS}&  \makecell[c]{0.0\%}  &\makecell[c]{51.63}   &     \makecell[c]{ 13.7\%}    &  \makecell[c]{ 41.8\%} \\
%   \makecell[c]{MASIA} &  \makecell[c]{98.6\%} & \makecell[c]{32.52}  & \makecell[c]{46.5\%}  &  \makecell[c]{ 28.5\%} \\\hline
%   \makecell[c]{M2I2(Ours)} & \makecell[c]{\textbf{165.6\%}}& \makecell[c]{\textbf{55.76}} & \makecell[c]{\textbf{98.7\%}}  & \makecell[c]{\textbf{59.3\%}}  \\
%   \bottomrule
% \end{tabular}
% \caption{Communication Efficiency}
% \label{table:efficiency}
% \end{table*}
% \subsection{Ablation}

%依次验证方法中提出的每个模块的有效性。初步setup:第一个图从总体角度验证大的模块的有效性，三条曲线分别是qmix(原始的baseline 没有加任何通信的模块)，qmix+mae+inverse(加上接收方的两个辅助任务，验证其有效性)，qmix+mae+inverse+dimensional gating（进一步验证gating的有效性）。后面是更加细粒度的验证，第一个图单独验证mae和inverse的效果，四条曲线,qmix, qmix+mae, qmix+inverse, qmix+mae+inverse。第二个图验证dim gating和obs gating的效果,三条曲线分别是ours without gating ours, with dim gating, ours with obs gating.

\subsection{Ablation (RQ2)}

\begin{figure}
    \centering
    \includegraphics[width=1.0\linewidth]{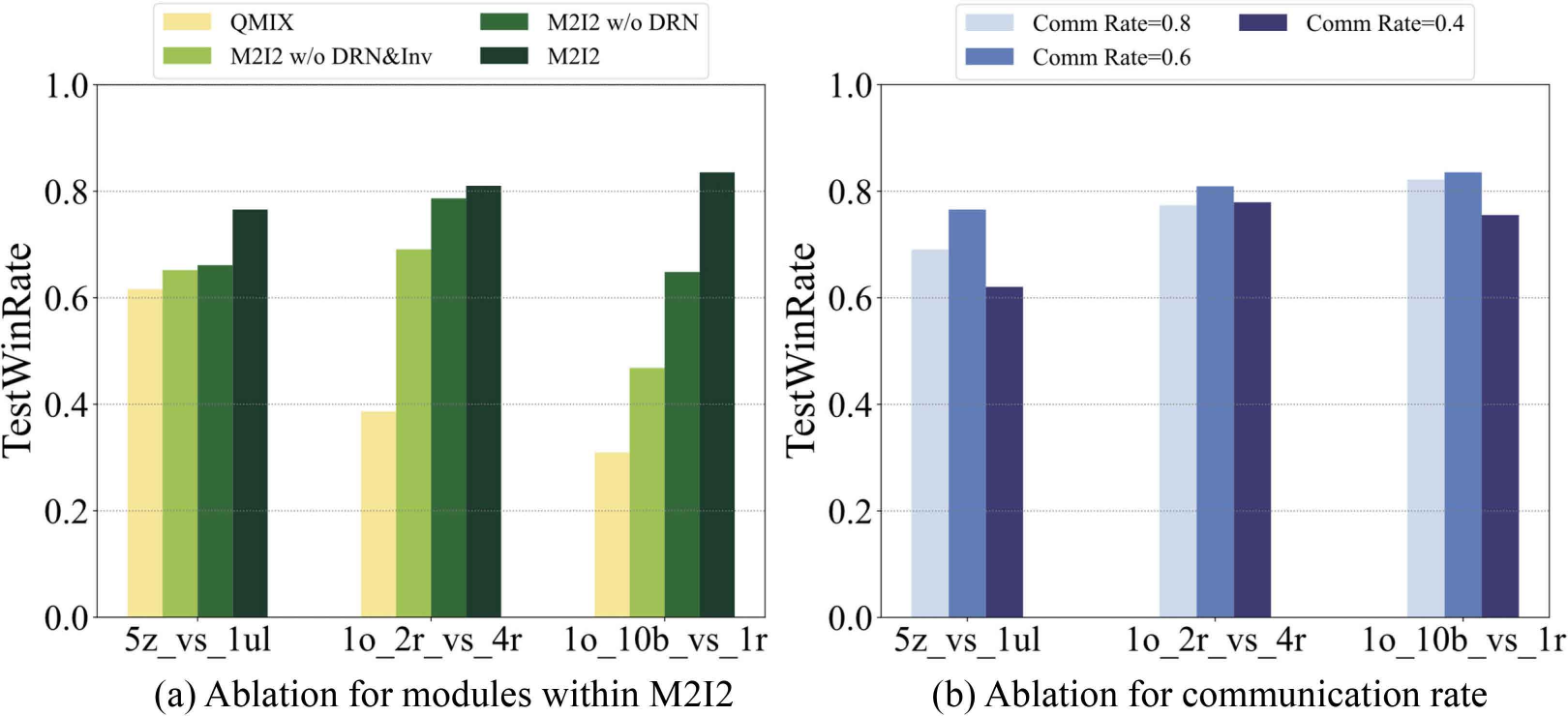}
    \caption{Ablation.}
    \label{ablation}
\end{figure}

In order to understand the contribution of each module within M2I2, we conduct an ablation study across three SMAC-Communication scenarios. The evaluated configurations are as follows: \textbf{M2I2} is our comprehensive method as proposed in the study. \textbf{QMIX} acts as a baseline, representing the fundamental functionality devoid of M2I2's enhancements. \textbf{M2I2 w/o DRN} is a variant of M2I2 operates without DRN and top-k filter mechanism. Instead, the observation level masking (i.e. when to communicate) is generated randomly during both the training and sampling processes. \textbf{M2I2 w/o DRN \& INV} is a further simplified version of M2I2, excluding both the DRN and inverse loss, retaining only the mask state modeling.  The results, as depicted in Figure \ref{ablation}(a), highlight the consistent performance improvements attributable to each component across the three scenarios. This underscores the effectiveness of our proposed auxiliary tasks and the DRN in enhancing the overall functionality of M2I2. 
The clear distinction in performance between these configurations serves to validate the integral role each module plays in the efficacy of the M2I2 framework. %Concretely, the impact of proposed modules is varied observed in Figure  \ref{ablation}(a)is attributed to the inherent differences in task characteristics. Specifically, the 5z\_vs\_1ul scenario is a homogenous task where even a communication-free QMIX can achieve commendable performance, leading to relatively modest gains from our contributions. In contrast, the 1o\_2r\_vs\_4r and 1o\_10b\_vs\_1r scenarios are heterogeneous tasks involving both scouting (observer entities) and combat units, necessitating higher communication precision, thereby accentuating the benefits of our proposed self-supervised objectives. Additionally, in 1o\_10b\_vs\_1r, where the larger number of entities intensifies the demands on strategic communication timing, thereby amplifying the advantages of DRN.

Furthermore, to gain insight into how varying communication frequencies impact M2I2's performance, we executed an ablation study with communication rates set at 0.8, 0.6, 0.4. The findings, depicted in Figure \ref{ablation}(b), consistently show the best performance at a communication rate of 0.6. This result suggests that an excessively high communication rate can introduce redundant and misleading information, whereas too low a rate may lead to critical information being overlooked. Intriguingly, reducing information by 0.6 using the meta mask still outperforms a random mask reduction of 0.4, underscoring the meta mask's proficiency in discerning and prioritizing key information. This delicate balance achieved by M2I2, efficiently filtering out non-essential data while preserving crucial information, significantly enhances the overall communication efficiency in complex multi-agent environments.

At last, to gain deeper insights into the selectively masking process utilized by M2I2, we conducted visualizations to identify which observations are most frequently masked. Additionally, we utilized t-SNE \cite{tsne} to project the learned representation vectors onto a two-dimensional plane. For a comprehensive presentation of these findings, %including detailed statistics on masking frequencies and t-SNE results, as well as other visualizations like the learning curves for $\mathcal{L}_{RC}$ and $\mathcal{L}_{INV}$, 
please refer to \textbf{Appendix} \ref{app:visualization}. 

% \begin{figure}[t]
%   \centering
%   \includegraphics[width=0.95\textwidth]{ablation.pdf}
%   \caption{Ablation.}
%   \label{ablation}
% \end{figure}

%\subsection{Visualization (RQ2)}
%Interestingly, we observed a notable capability of M2I2 to selectively share important observations. In SMAC, the observations are categorized into four key areas: agent movement features indicating possible movement directions, enemy features, ally features, and the agent's own unit characteristics. Significantly, enemy and ally features register as non-zero only when they are within an agent's range of observation. The intricate terrain further complicates the identification of enemies and allies.  
%Our analysis revealed a distinct pattern: The frequency of sharing enemy and ally features significantly increases when these features are observable by the agents. Such findings underscore M2I2's ability to discern and prioritize essential information, effectively filtering out irrelevant data. Additionally, we employed t-SNE \cite{tsne} to project the learned representation vectors into a two-dimensional plane. This analysis revealed that M2I2's representations possess robust discriminability, further showcasing the model's nuanced understanding of the environment. For an in-depth view of these findings, including detailed t-SNE results and other visualizations such as learning curves of $\mathcal{L}_{RC}$ and $\mathcal{L}_{INV}$, please refer to \textbf{Appendix} \ref{app:visualization}. 

\begin{figure}
    \centering
    \includegraphics[width=1.0\linewidth]{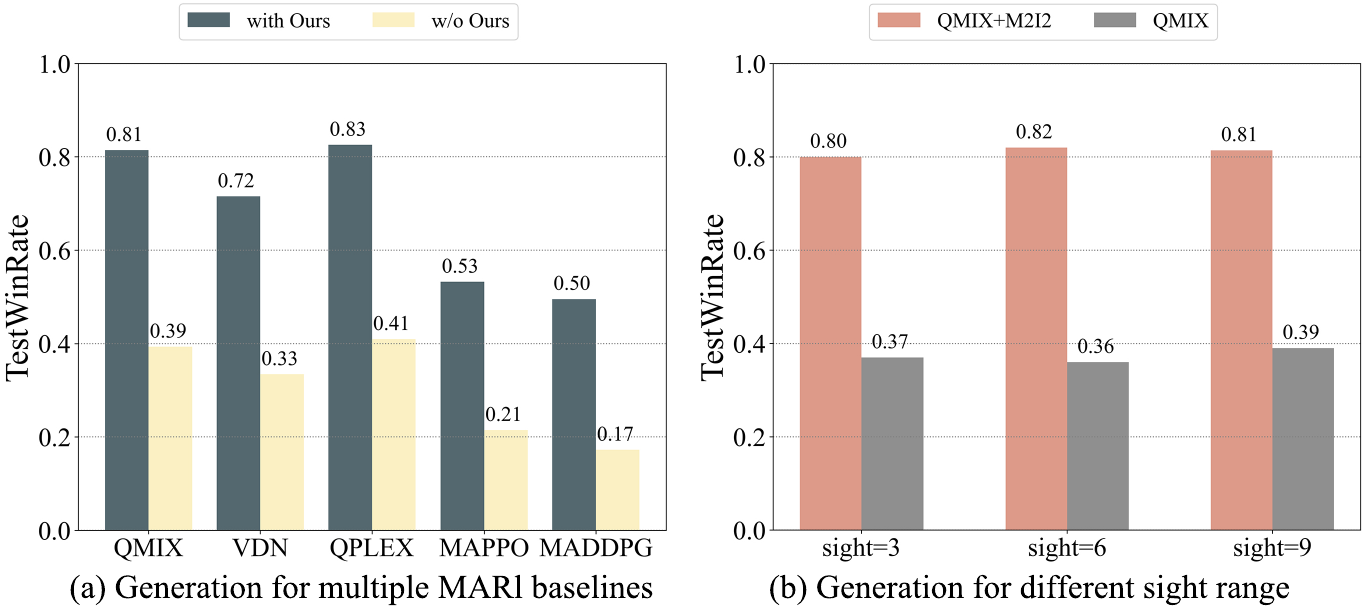}
    \caption{Generation.}
    \label{generation}
\end{figure}

\subsection{Generation (RQ3)}
%section\ref{performance} has demonstrate our method can scale to diffent tasks, different scenarios, different difficulty . In this section, we further test whether our method can scale to different baselines and whether it can work with different sight range.->两个图 第一个图曲线或者柱状图(vdn, qmix, qplex, mappo,  vdn+ours, qmix+ours, qplex+ours, mappo + ours) 第二个图曲线图（ours with sightrange=3, ours with sightrange=6, ourswithsightrange=9, qmix with sightrange=3, qmix with sight range=6, qmix with sightrange=9）
Our previous experiments have conclusively shown M2I2's robust performance in a variety of environments, encompassing scenarios with diverse complexities and scales. Building on this, we extend our evaluation of M2I2 to assess its generality across various MARL baselines, including QMIX, VDN, QPLEX, MAPPO and MADDPG. To provide a clear representation, we present the test win rates for the scenario $1o\_2r\_vs\_4r$ in Figure \ref{generation}(a). 
Remarkably, M2I2 demonstrates consistently superior performance across all these baselines, often achieving a significant margin of improvement. This observation shows that M2I2 is effective not only with off-policy algorithms but also with on-policy approaches, and not only with Q-learning methods but also with policy gradient methods, demonstrating M2I2’s broad applicability and effectiveness within the MARL domain. Additionally, we extend M2I2's application to scenarios featuring varying sight ranges. The results, as depicted in Figure \ref{generation}(b), confirm that M2I2 not only adapts well but also maintains consistent performance improvements across different sight ranges, further underscoring its versatility and efficacy in enhancing MARL strategies.
% \begin{figure}[t]
%   \centering
%   \includegraphics[width=0.95\textwidth]{generation.pdf}
%   \caption{Generation.}
%   \label{generation}
% \end{figure}

\section{Conclusion}
\label{conclusion_limitation}
In this work, we delve into the complexities of multi-agent information integration. We introduce M2I2, an approach that incorporates two auxiliary tasks to enhance communication efficiency. We specifically design a MAE and an inverse model. These elements play a crucial role in guiding the processes of information filtering and integration, thereby significantly enhancing the agents' ability to navigate uncertain environments and dynamically adapt to their teammates. To substantiate our claims, we conduct exhaustive experiments across a multitude of benchmarks. The results from these tests not only validate the effectiveness of M2I2 but also its efficiency and adaptability in various multi-agent scenarios. %These findings highlight M2I2's potential to significantly advance the field of multi-agent communication. 

% One limitation of M2I2 is the incomplete results visualization. In future work, we plan to delve deeper into which observations are being masked by M2I2 and what kind of representations the model is learning. This enhanced understanding will help further refine the model's application and effectiveness.

\bibliography{aaai25}

\newpage
\appendix
\onecolumn

\section{Notations}
\label{app:notations}
For clarity and precise symbol definitions, we provide a comprehensive list of notations used in this work in Table. \ref{tab:notations}.

\begin{table}[h]
\centering
\begin{tabular}{ll}
  \toprule
  \textbf{Notations}  & \\
  \midrule
  $N$         & a collective of agents \\
  $n$         & number of agents \\
  $S$         & a set of global states \\
  $O$         & accessible local observations \\
  $A$         & available actions \\
  $\mathbb{O}$     & observation function \\
  $P$         & transition function \\
  $R$         & reward function \\
  $\gamma$       & discount factor \\
  $M$         & the set of messages \\
  $s_t$        & global state at time-step $t$ \\ 
  $o_i^t$       & observation of agent $i$ at time-step $t$ \\
  $m_i^t$       & message sent by agent $i$ at time-step $t$ \\
  $z_i^t$       & integrated information by agent $i$ at time-step $t$ \\
  $a_i^t$       & selected action of agent $i$ at time-step $t$ \\
  $a_t$         & joint action of all agents at time-step $t$ \\
  $s_t^{\prime}$  & reconstructed state at time-step $t$\\
  $a_t^{\prime}$  & predicted joint-action at time-step $t$\\
  $\pi(o_i^t, c_i^t)$ & local policy \\
  $\mathcal{\omega}_i^t$       & the importance of agent $i$'s observed information at time-step $t$\\
  $Q_{tot}$     & total q-value\\
  $\mathcal{L}_{RL}$   & loss of RL\\
  $\mathcal{L}_{RC}$  & reconstructed loss\\
  $\mathcal{L}_{INV}$   & inverse loss\\
  $\mathcal{L}_{M2I2}$ & M2I2 loss \\
  $\theta_{E}$  & parameters of message encoder\\
  $\theta_{\pi}$  & parameters of policy network\\
  $\theta_{D}$  & parameters of state decoder\\
  $\theta_{I}$  & parameters of inverse model\\
  $\theta_{DRN}$ & parameters of DRN\\
  $\theta_{trial}$ & trial parameters for meta learning\\
  
  \bottomrule
\end{tabular}
\caption{Notations}
\label{tab:notations}
\end{table}

\begin{algorithm}[htbp]
\begin{algorithmic}
    \STATE Initialize replay buffer $D$, 
    \STATE Initialize the random parameters of Dimensional Rational Network $\theta_{DRN}$, Message Encoder $\theta_{E}$, State Decoder $\theta_{D}$, Inverse Model $\theta_{I}$ and  Policy Network $\theta_{\pi}$
    %\STATE Initialize $\theta$, the parameters for Policy $\pi$; $\phi$, the parameters for Selective Engagement
    \STATE Set learning rate $\alpha$ and max training episode $E$
    \FOR {episode in $1,...,E$} 
        \FOR{each agent $i$ }
            \STATE \textbf{Sending Phase:} 
            \STATE Encode importance weight $\mathcal{\omega}_i$ from observation $o_i^t$
            \STATE Compute the masked importance weight $\mathcal{\omega}_i^{\prime}$ with Top-k Scheduler by Equation \ref{topk}
            \STATE Generate the shared message $m_i^t$ with $\mathcal{\omega}_i^{\prime}$ and $o_i^t$ by Equation \ref{message_generate}
            \STATE \textbf{Receiving Phase:}
            \STATE Encode the integrated representation $z_i^t$ from the received message by Equation \ref{intergrated_representation}
            \STATE Make cooperative decisions by Policy Network for $a_i^t$ by Equation \ref{action_select}
        \ENDFOR
        \STATE Store the trajectory in replay buffer D
        \STATE Sample a minibatch of trajectories from D
        % \STATE Update observation encoder, evidence encoder and policy network using loss function denoted in \cite{qmix}
        \STATE Compute the reconstructed loss $\mathcal{L}_{RC}$ by Equation \ref{lossRC}
        \STATE Compute the inverse loss $\mathcal{L}_{INV}$ by Equation \ref{lossINV}
        \STATE Compute the reinforcement Loss $\mathcal{L}_{RL}$
        \STATE Update the combined set of parameters $\mathcal{\theta} = (\theta_{E}, \theta_{D} , \theta_{I} , \theta_{\pi} )$ by Equation \ref{lossM2I2}
        \STATE Compute the trail weight $\mathcal{\theta}_{trail}$ after one gradient update of $\mathcal{\theta}$ using the M2I2 loss by Equation \ref{trailcompute}
        \STATE Update the parameters of Dimensional Rational Network $\theta_{DRN}$ by Equation \ref{updatemeta}
    \ENDFOR
\end{algorithmic}
\caption{$\text{M2I2}$}
\label{alg:M2I2full}
\end{algorithm}

\section{Implementation Details}
\label{app:implementation}
To explain the communication process and training paradigm, we provide the pseudo-code for M2I2 in Algorithm \ref{alg:M2I2full}.

\begin{figure*}[h]
  \centering
  \includegraphics[width=0.7\textwidth]{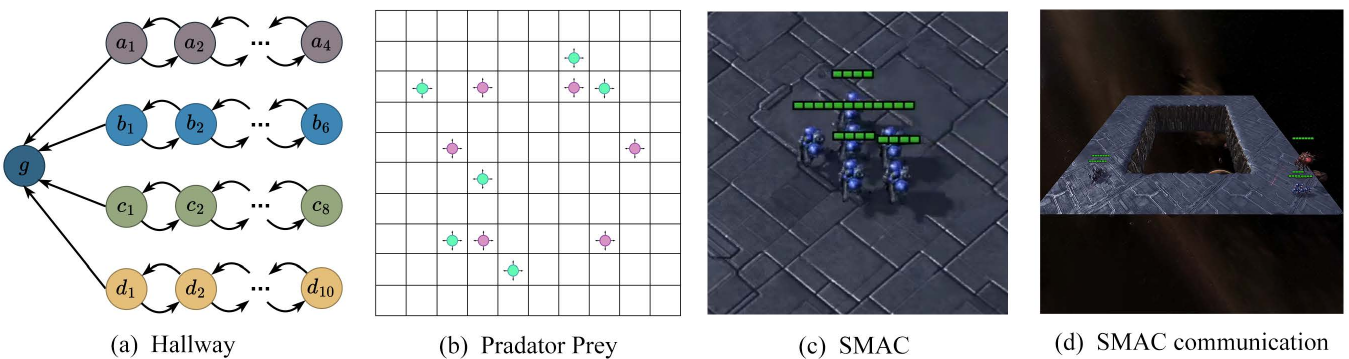}
  \caption{Multiple environments considered in our experiments.
  }
  \label{senarios}
\end{figure*}

\section{Experimental Setup}
\label{app:experimentalsetup}
All the Result are reported by averaging the results of 5 random seeds. The four test environments of the experiment are shown in Figure \ref{senarios}.

\subsection{Environment}
\textbf{Hallway}. This task revolves around multiple Markov chains where $n$ are initially distributed randomly across $n$ chains with varying lengths. The goal is for all agents to simultaneously reach the goal state, despite the constraint of partial observability. For increased challenge, $n$ is set to 4, and each chain has a unique length specified as (4, 6, 8, 10). At each time step, an agent has a limited observation of its current position, and it can choose from three actions: move left, move right, or remain stationary. An episode concludes when any agent reaches state $g$. The agents collectively succeed and receive a shared reward of 1 only if they all reach state $g$ concurrently. Otherwise, the reward is 0.

\textbf{Hallwaygroup}. This is a variant of Hallway which agents are divided into different groups and different groups have to arrive at different times. To intensify the challenge, we escalated the complexity by increasing both the number of agents and the lengths of the Markov chains. Specifically, in the Hallway benchmark, we set the number of agents to 4, with Markov chain lengths varying as (4,6,8,10). In the Hallwaygroup variant, we increased the number of agents to 7, dividing them into two groups. The lengths of the Markov chains for these two groups were set to (3,5,7) and (4,6,8,10), respectively.

\textbf{Predator Prey (PP)}.
In this task, the objective for predators is to capture randomly moving preys. Each predator has the ability to navigate in four distinct directions, but their perspectives are limited to local views. The game dynamics involve multiple predators attempting to capture the same prey simultaneously, resulting in a team reward of 1. However, if only a single predator successfully captures a prey, they incur a penalty with a score of -2. The game concludes when all preys have been captured. To introduce varying difficulty levels, different grid sizes and numbers of agents are employed for the Predator-Prey (PP) scenarios. The specific configurations for two PP scenarios are detailed in Table. \ref{tab:pp}.

\begin{table}[h]
\centering
  \caption{The configurations of  PP scenarios}
  \begin{tabular}{cccc}\toprule
    \textit{} & \textit{Grid size} & \textit{$n_{predators}$} & \textit{$n_{preys}$} \\ \midrule
    Medium & 10 * 10 & 6 & 6   \\
    Hard & 15  * 15 & 8 &  8\\ 
    \bottomrule
  \end{tabular}
\label{tab:pp}
\end{table}

\textbf{StarCraft Multi-Agent Challenge (SMAC)}.
This task revolves around a series of complex scenarios inspired by StarCraft\uppercase\expandafter{\romannumeral2}, a real-time strategy game. Decentralized agents engage in combat against the built-in AI, each having a limited field of vision restricted to adjacent units. Observations include relative positions, distances, unit types, and health statuses. Agents struggle to perceive the status of entities beyond their immediate vicinity, creating uncertainty. The action space varies across scenarios, often including move, attack, stop, and no-option. During training, global states with coordinates and features of all agents are accessible. Rewards are based on factors like damage infliction, eliminating units, or victory.

\textbf{SMAC-Communication}. To emphasize the role of communication, we select three super hard maps and further adopt the configuration used in \cite{ndq}. 
The specifics of the chosen scenarios are delineated as follows. 
%The specifics of the chosen scenarios are detailed in Table. \ref{tab:smac}. Furthermore, we provide the delineated descriptions of SMAC-Communication maps as follows.

\textbf{$5z\_vs\_1ul$}. A team of 5 Zealots faces a formidable Ultralisk. Victory requires mastering a complex micro-management technique involving positioning and attack timing.

\textbf{$1o\_10b\_vs\_1r$}. In a cliff-dominated terrain, an Overseer spots a Roach. 10 Banelings aim to eliminate the Roach for victory. Banelings can choose silence, relying on the Overseer to communicate its location, testing communication strategy efficiency.

\textbf{$1o\_2r\_vs\_4r$}. An Overseer encounters 4 Reapers. Allied units, 2 Roaches, must locate and eliminate the Reapers. Only the Overseer knows the Reapers' location, requiring effective communication for success.

\subsection{Network architecture}
%   Table \ref{tab:netarch} reports the network architecture of M2I2.
\begin{table}[h]
\centering
\begin{tabular}{ll}
  \toprule
  \textbf{Module}  & Architecture \\
  \midrule
  \multirow{1}*{Message Encoder}      & Q:Linear(obs\_dim,                                         16)\\
                                      &K:Linear(obs\_dim, 16)\\
                                      &V:Linear(obs\_dim, 32)\\ 
                                     & RNN(32, 32)\\
                                     %& N*Linear(32, 32)\\
                                     &Linear(32,8*n\_agent)\\\hline
  State Decoder                 & Linear(8*n\_agent,32)\\
                                %&N*Linear(32,32)\\
                                &Linear(32,state\_dim)\\\hline
  Inverse Model                 & 
                                Linear(8*n\_agent,64)\\
                                %& ReLU,\\
                                %& Linear(64,64)\\
                                &
                                Linear(128,64)\\
                                %& ReLU,\\
                                &Linear(64,n\_agent*n\_action)\\
    \bottomrule
  Policy Network                & Linear(obs\_dim,32)\\
                                & Linear(32+8*n\_agent,32)\\
                                & RNN(32, 32)\\
                                & Linear(32, n\_action)\\
  
  \bottomrule
       DRN                      &Linear(obs\_dim,32)\\
                                &RNN(32,32)\\
                                &Linear(32,obs\_dim)\\
\bottomrule
\end{tabular}
\caption{Network architecture of M2I2}
\label{tab:netarch}
\end{table}

\subsection{Hyper-parameters}
%Table \ref{tab:hyparam} shows the hyper-parameters of M2I2.
\begin{table}[h]
\centering
\begin{tabular}{ll}
  \toprule
  \textbf{Hyper-Parameters}  & \\
  \midrule
  epsilon start               & 1.0 \\
  epsilon finish             & 0.05 \\
  epsilon anneal time        & 50000 \\
  buffer size                & 5000 \\
  target update interval     & 200 \\
  hidden dimension for mixing network           & 32 \\
  $\beta$                 & 1 \\
  mask ratio  & 0.4 \\

  \bottomrule
\end{tabular}
\caption{Hyper-Parameters of M2I2}
\label{tab:hyparam}
\end{table}

\section{Visualization}
\label{app:visualization}
% To showcase the effectiveness of each module within M2I2, we conduct visualizations to illustrate what features does DRN focus on more, which pieces of observations are masked and what the integrated representation has learned. Initially, to visualize what kind of knowledge the DRN has learned, we performed a statistical analysis of communication frequencies based on the dimensions of observations. In SMAC, information outside the field of view is padded with zeros in the observation. We found that the communication frequency of effective ally and enemy information within the field of view is significantly higher than the communication frequency of ineffective information outside the field of view Figure \ref{Feature Communication}. In some scenarios with complex terrain where it is difficult to detect enemies, the communication frequency of observed enemy information is notably higher than that of observed ally information. This indicates that the DRN can recognize the importance of observed information for decision-making and auxiliary tasks, performing selective information sharing and masking. 

To showcase the effectiveness of each module within M2I2, we conduct visualizations to illustrate what features does DRN focus on more, which pieces of observations are masked and what the integrated representation has learned.

\begin{figure}[htp]
    \centering
    \includegraphics[width=1.0\linewidth]{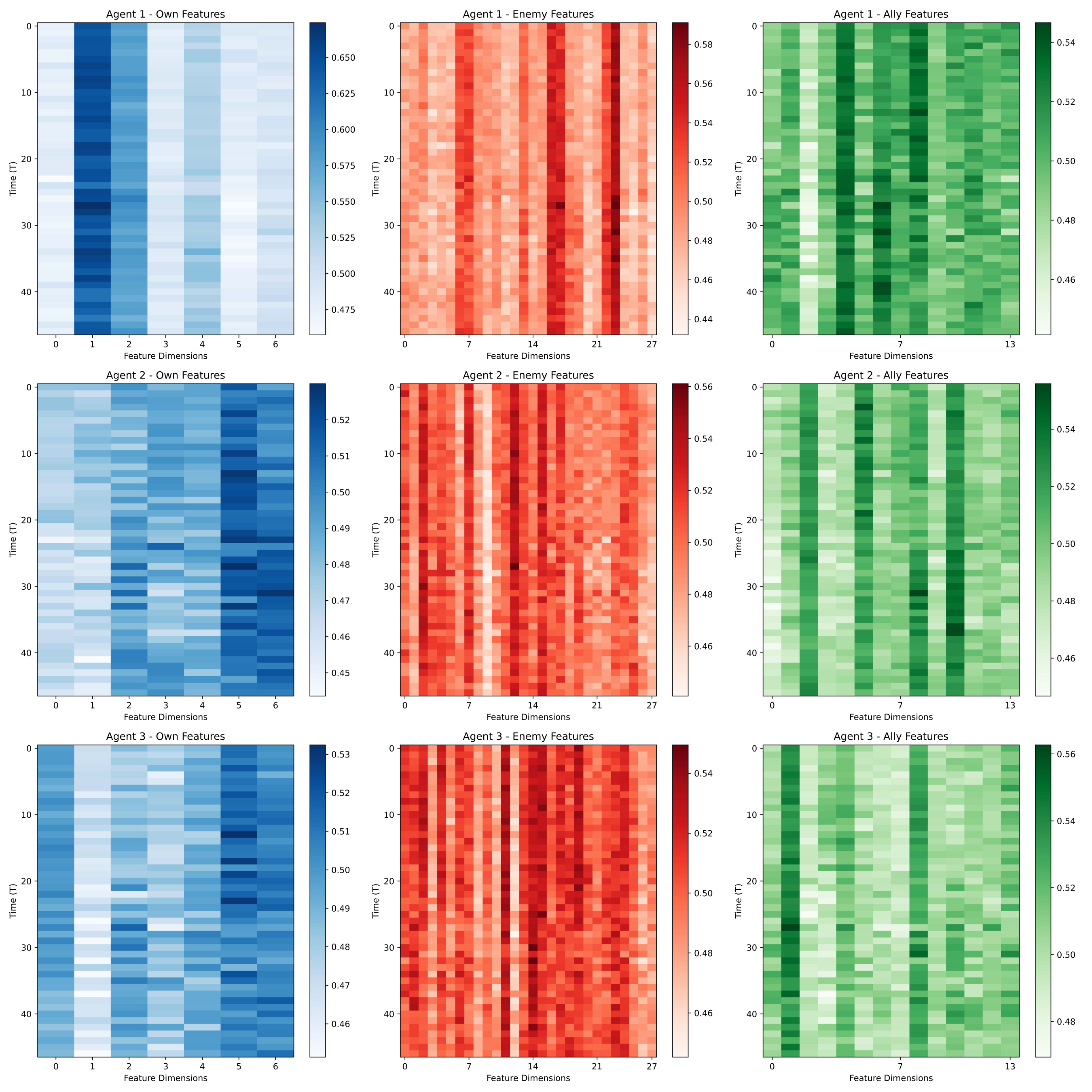}
    \caption{Visualization of weight learn by DRN}
    \label{meta_feature_weight}
\end{figure}

\subsection{Visualization of weight learen by DRN}
%To evaluate the adaptive capability of DRN, we visualized the learned importance of each observation component and found that DRN can effectively allocate varying levels of importance across different agent types and observation types. Moreover, these importance weights adapt dynamically as time and situations evolve. The detailed analysis can be summarized into three key points:
%\begin{itemize}
 %   \item \textbf{Agent level analysis.} Overall, the distribution of importance for each agent 1 is very different from agent 2 and 3, as in this task, Agent 1 is primarily responsible for observing enemy information, while Agents 2 and 3 are tasked with moving towards the enemy’s location then destroy the them.
   % \item \textbf{Observation level analysis.}
  %  \item \textbf{Time level analysis.}
%\end{itemize}

To evaluate the adaptive capability of DRN, we visualized the learned importance of each observation component in Fig. \ref{meta_feature_weight} and found that DRN can effectively allocate varying levels of importance across different agent types and observation types. Moreover, these importance weights adapt dynamically as time and situations evolve. The detailed analysis can be divided into three-fold:
\begin{itemize}
    \item \textbf{Agent level analysis.} In the task we evaluated, Agent 1 primarily focuses on observing enemy information, while Agents 2 and 3 are tasked with moving towards the enemy’s location to destroy them. The visualization reveals that the $\mathcal{\omega}_i^t$ matrices for Agents 2 and 3 are more similar to each other and less similar to the matrix for Agent 1. This indicates that DRN differentiates between the significance of various agent types.
    \item \textbf{Observation level analysis.} We found that DRN assigns different levels of importance to various observation dimensions, highlighting its ability to prioritize critical information effectively.
    \item \textbf{Time level analysis.} Lastly, we observed that the importance of certain observations dynamically changes over time, further demonstrating DRN’s adaptive capability in response to evolving situations.
\end{itemize}

%Concretely, from the \textbf{agent level}, we loaded the M2I2 model that was trained for 200 million steps on the $1o\_2r\_vs\_4r$ map and recorded the $\mathcal{\omega}_i^t$  values after each time step's observation was evaluated by the $DRN$. In this task, Agent 1 is primarily responsible for observing enemy information, while Agents 2 and 3 are tasked with moving towards the enemy’s location then destroy the them. From the visualization, we can observe that the $\mathcal{\omega}_i^t$ matrices for Agents 2 and 3 are more similar to each other, while they are less similar to the matrix for Agent 1.This indicates that the $DRN$ differentiates between various levels of significance among different agent types. Furthermore, from the \textbf{observation level}, we find that DRN can allocate different importance for different observation dimensions.  At last, from the \textbf{time level}, we observe that the importance of some observations are dynamically changing over time, which further demonstrating the adaptive ability of DRN. 

\subsection{Visualization of mask probability}
\begin{figure}[htp]
    \centering
    \includegraphics[width=1.0\linewidth]{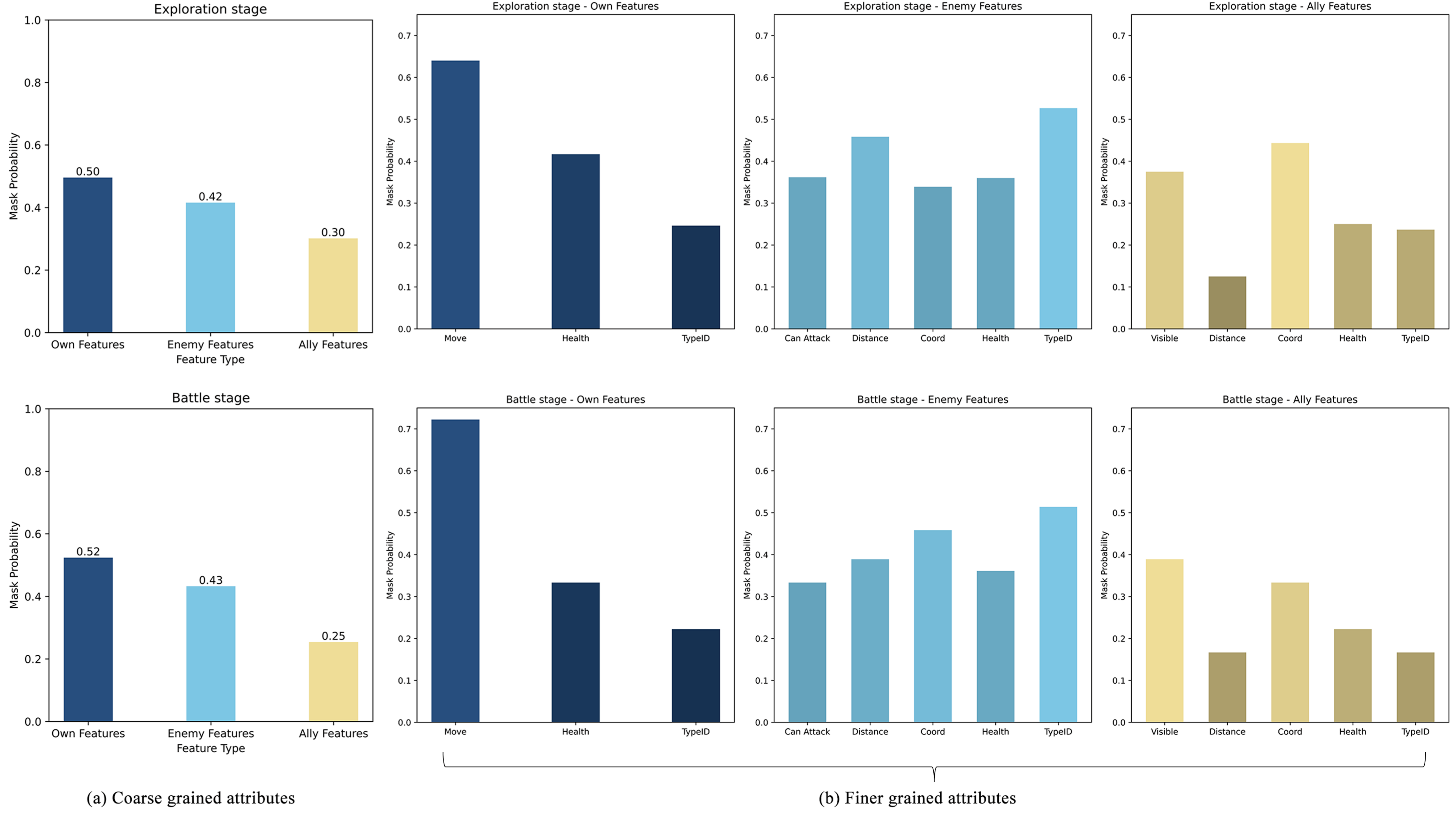}
    \caption{(a) Presents the probability of masking for the three types of observations at different stages. (b) Illustrates the probability of masking for each specific attribute within each observation type across various stages.}
    \label{Attribute_visualization}
\end{figure}

To gain deeper insights into the meta-learned mask generating policies (i.e., communication policies), we present a visualization of the mask ratio for different observations in Figure \ref{Attribute_visualization}. Specifically, we divided each episode into two stages: the first 80\% of time steps are categorized as the Explore stage, and the remaining 20\% as the Battle stage. During the Explore stage, the agent primarily focuses on moving and sharing information about enemy locations, with an emphasis on exploration and navigation. In the Battle stage, the agent identifies the enemy and engages in combat. %Based on these stages, we conducted a finer-grained analysis of the time dimension of the weight matrix $\mathcal{\omega}$ obtained from DRN. We counted the probability of each attribute being masked in the observations. 

At a high level, we categorize the observations into three types: observations of the agent itself, observations of enemy agents, and observations of ally agents. As shown in Figure \ref{Attribute_visualization}(a), we observe that DRN considers ally features to be more critical for decision-making in both stages, leading to a lower proportion of these features being masked compared to other types of observations.

To gain deeper insights, we further categorized the own, enemy, and ally features into more specific attributes. As depicted in Figure \ref{Attribute_visualization}(b), we observed that DRN's masking tendencies shift between the stages depending on the attribute. During the Exploration stage, the agent relies heavily on enemy information for positioning and movement, resulting in the lowest masking rate for enemy location attributes. However, in the Battle stage, as movement becomes less critical, the importance of enemy health information increases, leading to a higher retention of health-related attributes.

Overall, this visualization highlights that DRN dynamically adjusts the importance weights of different observation attributes over time to align with the evolving demands of the task.

\subsection{Visualization of learned representations}
At last, to visualize what kind of representations the encoder has learned,  we use t-SNE \cite{tsne} to project the learned representation vectors into a two-dimensional plane. As illustrated in Figure \ref{meta_feature_weight}, we conduct a visualization analysis on the map $1o\_2r\_vs\_4r$. To distinguish aggregated representations at different time-steps, we mark larger time-steps with darker shades of the dots. We observe that the aggregated representations can be well distinguished by phases. Projected representations in the seeking phase are far from those in the battle phase, demonstrating that M2I2's representations exhibit robust discriminability, indicating the model's nuanced understanding of the environment.

\begin{figure}[htp]
   \centering
   \includegraphics[width=0.5\textwidth]{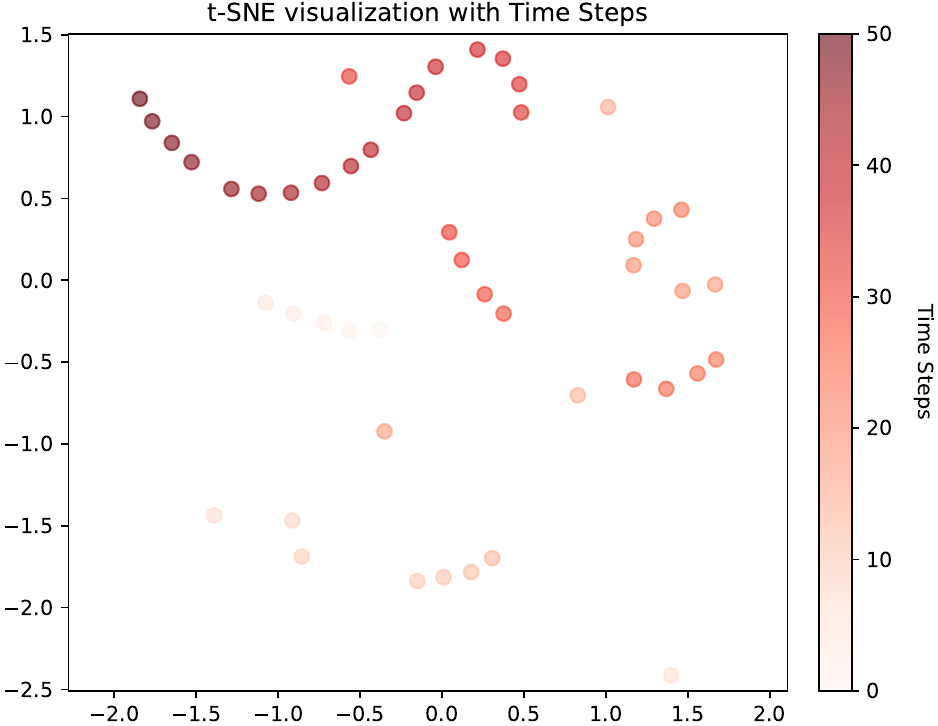}
   \caption{Visualization of learned representations
   }
   \label{Visualization}
 \end{figure}

\subsection{Learning curves for self-supervised learning loss}
\begin{figure}[htp]
   \centering
   \includegraphics[width=0.9\textwidth]{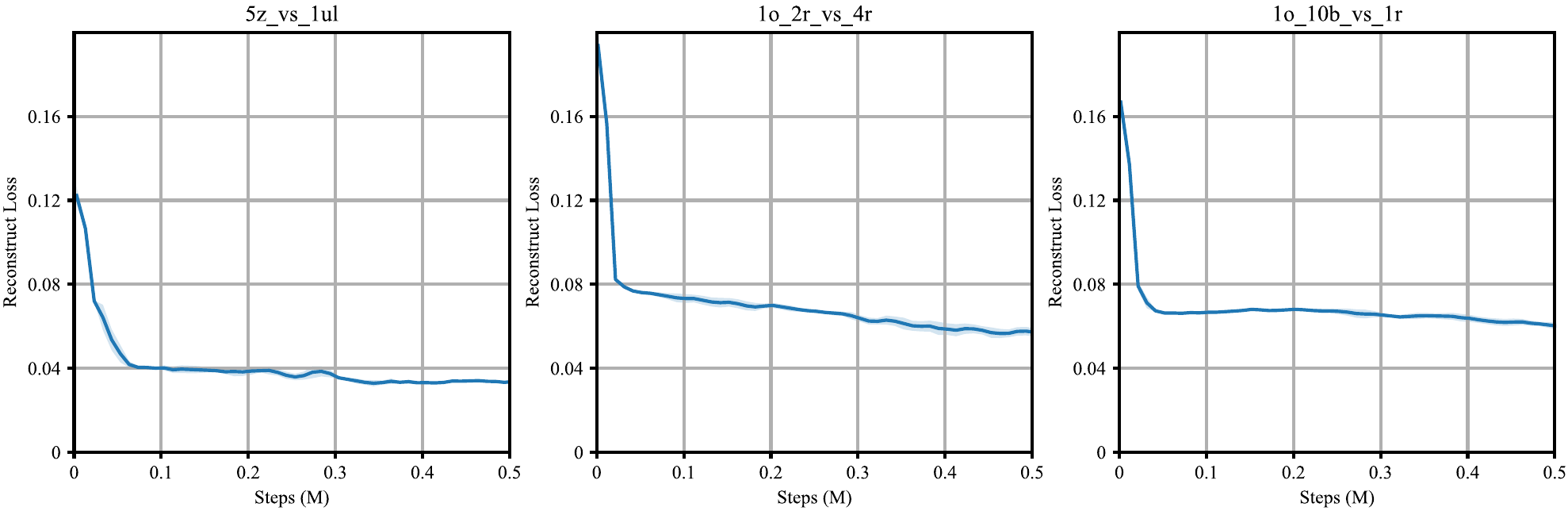}
   \caption{Reconstruct Loss
   }
   \label{RC Loss}
 \end{figure}

\begin{figure}[htp]
   \centering
   \includegraphics[width=0.9\textwidth]{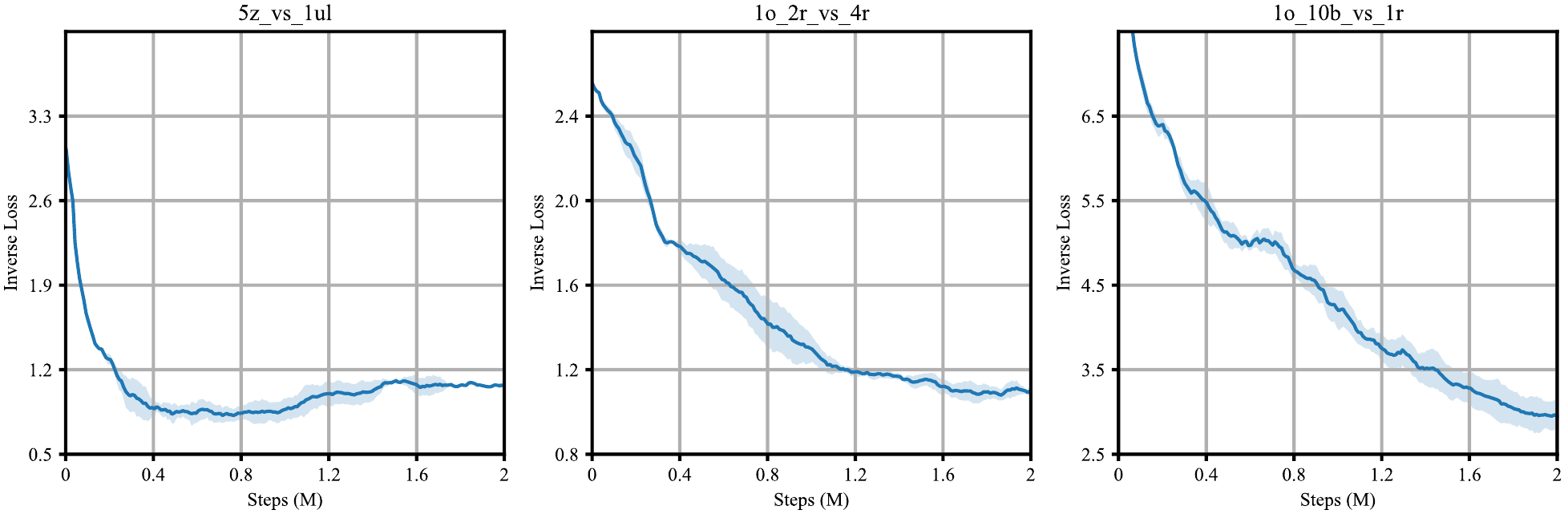}
   \caption{Inverse Loss
   }
   \label{Inverse Loss}
\end{figure}

In addition to the visual representation, we provide learning curves for $\mathcal{L}_{RC}$ and $\mathcal{L}_{INV}$ in Figure \ref{RC Loss} and Figure \ref{Inverse Loss}, respectively. The observed quick convergence of $\mathcal{L}_{RC}$ and $\mathcal{L}_{INV}$ suggests that our masked modeling and inverse model indeed equip agents with the ability to reconstruct global states and infer joint actions based on limited communication resources. These curves offer insights into the training dynamics and convergence of the reconstruction and inverse modeling tasks.

% \begin{figure}[h]
%    \centering
%    \includegraphics[width=0.5\textwidth]{Mask ratio.pdf}
%    \caption{Feature Communication
%    }
%    \label{Feature Communication}
%  \end{figure}

\section{Details of Computational Resources}
\label{app:compute}
The computational experiments described in this paper were executed on a dedicated high-performance computing cluster to ensure the reproducibility and efficiency of the results. Below, we provide the detail of the computational resources used:

\begin{itemize}
    \item \textbf{GPU Specifications:}
    \begin{itemize}
        \item Quantity: 3 NVIDIA TITAN Xps
        \item Memory: 12 GB GDDR6X per GPU
    \end{itemize}
    
    \item \textbf{CPU Specifications:}
    \begin{itemize}
        \item Model: Intel(R) Xeon(R) Silver 4114 CPU 
        \item Architecture: x86\_64
        \item Base Clock Speed: 2.20GHz
    \end{itemize}
    
    \item \textbf{Software and Frameworks:}
    \begin{itemize}
        \item Operating System: 16.04.1-Ubuntu SMP
        \item Machine Learning Libraries: torch 2.1.0
    \end{itemize}
\end{itemize}

In Table.\ref{tab:compute parameters} we present the number of models' parameters and running time of our baselines on the $1o\_2r\_vs\_4r$. In the Experiment, M2I2's parameters and training time, 0.82-1.19 and 0.65-1.47 times those of baselines respectively, which indicates its computational efficiency.

\begin{table}[h]
    \centering
    \begin{tabular}{|c|c|c|c|c|c|c|} \hline 
         &  M2I2&  MAISA&  QMIX&  TarMAC &  MAIC& SMS\\ \hline 
         Parameters&  182,458&  178,348&  79,605&  123,899&  123,017& 84,326\\ \hline 
         Memory/Process&  368MIB&  358MIB&  309MIB&  330MIB&  325MIB& 808MIB\\ \hline 
         Time/episode&  0.45s&  0.4s&  0.082s&  0.15s&  0.44s& 0.9s\\ \hline
    \end{tabular}
    \caption{Parameters and running time of our baseline }
    \label{tab:compute parameters}
\end{table}

\clearpage

\iffalse
\begin{table*}[h]
  \caption{The detailed information of selected SMAC scenarios}
  \label{tab:smac}
  \begin{tabular}{cccc}\toprule
    \textit{} & \textit{Difficulty} & \textit{Allied units} & \textit{Enemy units} \\ \midrule
    3s5z\_vs\_3s6z & Super Hard & 3 Stalkers and 5 Zealots & 3 Stalkers and 6 Zealots   \\
    corridor & Super Hard & 6 Zealots &  24 Zerglings\\
    MMM2 & Super Hard & 1 Medivac, 2 Marauders and 7 Marines & 1 Medivac, 3 Marauders and 8 Marines  \\ 
    5z\_vs\_1ul & Communication Hard & 5 Zealots & 1 Ultralisk  \\
    1o\_10b\_vs\_1r & Communication Hard & 1 Overseer and 10 Banelings & 1 Roach  \\ 
    1o\_2r\_vs\_4r & Communication Hard & 1 Overseer and 4 Reapers & 2 Roaches  \\
    
    \bottomrule
  \end{tabular}
\end{table*}
\fi

% \bibliography{aaai25}

\end{document}